# Comment on "Self-Purification in Semiconductor Nanocrystals"

In a recent Letter [1], Dalpian and Chelikowsky claimed that formation energies of Mn impurities in CdSe nanocrystals (NCs) increase as the size of the NC decreases, and argued that this size dependence leads to "self-purification" of small NCs. They presented density-functional-theory (DFT) calculations showing a strong size dependence for Mn impurity formation energies, and proposed a general explanation. In this Comment we show that several different DFT codes, pseudopotentials, and exchange-correlation functionals give a markedly different result: We find no such size dependence. More generally, we argue that formation energies are not relevant to substitutional doping in most colloidally grown NCs.

We performed DFT calculations of the formation energies of Mn impurities in passivated CdSe NCs of different sizes, using methods similar to Ref. [1]. We computed the Mn impurity formation energy relative to its value in bulk CdSe, represented by a 216-atom supercell. We performed complete structural relaxation, using the projector-augmented-wave (PAW) method within VASP. We confirmed our results using ultrasoft pseudopotentials within VASP, and Troullier-Martins pseudopotentials with full $-Z/r$ potentials for passivation within ABINIT. Details can be found in Ref. [2].

Our results are shown in Fig. 1. For all NCs studied the Mn impurity formation energy differs by less than 0.03 eV from its value in bulk CdSe. Such small deviations from zero are not physically meaningful. In other words, we find no evidence from DFT for any dependence of impurity formation energy on NC size.

We turn now to understanding these results. When Mn substitutes for Cd in a CdSe NC, an occupied level is created in the gap. As the NC size decreases, quantum confinement shifts the host valence-band (VB) edge downward. Therefore, relative to the VB edge, the gap level shifts deeper into the gap. In Ref. [1] it was argued that this relative shift increases the formation energy of a Mn impurity. But this analysis is incomplete. The gap level arises from hybridization of Mn atomic $d$ and Se dangling-bond $p$ orbitals, and the bonding combination of these orbitals (the "resonant level") largely cancels the energy cost from the antibonding gap level. Hence the formation energy of a Mn impurity should remain approximately constant [2], as we find.

Reference [1] also reports that other defects in CdSe, such as Cd vacancies, show an increase in defect formation energy for small NCs. On this basis, a general conclusion about self-purification of impurities and defects in small NCs was proposed.

We believe this conclusion is unwarranted. At the temperatures used in colloidal growth, 350 °C and below, diffusion barriers for most defects cannot be readily overcome. Since thermal equilibrium cannot be reached, the formation energies of defects and impurities are not relevant to understanding their formation. In Ref. [3] we proposed a very different view in which the kinetics of impurity adsorption, as well as surfactant binding, plays the decisive role.

Moreover, even if thermal equilibrium were reached, size dependence of formation energies does not necessarily imply self-purification of small NCs. For example, in bulk CdSe the formation energy of a Cd vacancy is in the range 2.2 to 3.4 eV, depending on the Cd chemical potential [4]. This formation energy implies that the equilibrium concentration of Cd vacancies cannot exceed 1 part in $10^{18}$ at colloidal temperatures. Even if the formation energy increases in small NCs, this effect will be irrelevant since no vacancies would be expected in NCs of any size.

*Note added:* In their Reply [5] to this Comment, the authors of Ref. [1] have revised their DFT results as shown in Fig. 1. The revised results are essentially independent of nanocrystal size. The remaining offset from the bulk formation energy is about 0.1 eV. It is easy to demonstrate that an error of 0.1 eV can arise from unconverged sampling of the bulk supercell Brillouin zone. This would bring the results of Ref. [5] into excellent agreement with our own.

M.-H. Du,[1] S.C. Erwin,[1] Al.L. Efros,[1] and D.J. Norris[2]

[1] Center for Computational Materials Science
Naval Research Laboratory
Washington, D.C. 20375, USA

[2] Department of Chemical Engineering and Materials Science
University of Minnesota
Minneapolis, Minnesota 55455, USA

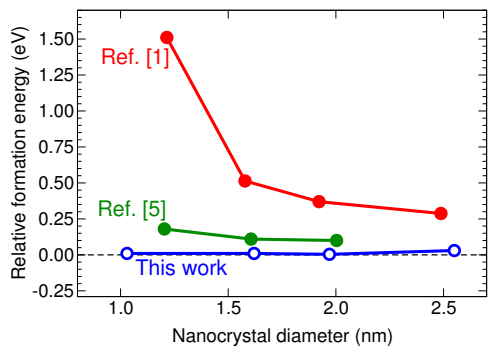

FIG. 1 (color online). Formation energy of a substitutional Mn impurity in a CdSe nanocrystal as a function of diameter.